\documentclass[conference]{IEEEtran}
\IEEEoverridecommandlockouts

\usepackage{cite}
\usepackage{amsmath,amssymb,amsfonts}
\usepackage{algorithmic}
\usepackage{graphicx}
\usepackage{textcomp}
\usepackage{xcolor}

\usepackage{algorithm}
\usepackage{array}
\usepackage{booktabs}
\usepackage{wrapfig}
\usepackage{stfloats}
\usepackage{url}
\usepackage{verbatim}
\usepackage{subcaption}
\usepackage{mathtools}
\usepackage{pgfplots}
\usepackage{tabularx}
\pgfplotsset{compat=1.16}
\usepackage[colorlinks=true, linkcolor=black, citecolor=black, urlcolor=blue]{hyperref}

\def\BibTeX{{\rm B\kern-.05em{\sc i\kern-.025em b}\kern-.08em
    T\kern-.1667em\lower.7ex\hbox{E}\kern-.125emX}}
\begin{document}

\title{HyperEF 2.0: Spectral Hypergraph Coarsening via Krylov Subspace Expansion and Resistance-based Local Clustering
}



\author{\IEEEauthorblockN{Hamed Sajadinia}
\IEEEauthorblockA{
\textit{Stevens Institute of Technology}\\
Hoboken, NJ, USA \\
hsajadin@stevens.edu}
\and
\IEEEauthorblockN{Zhuo Feng}
\IEEEauthorblockA{
\textit{Stevens Institute of Technology}\\
Hoboken, NJ, USA \\
zfeng12@stevens.edu}
}

\maketitle

\begin{abstract}
This paper introduces HyperEF 2.0, a scalable framework for spectral coarsening and clustering of large-scale hypergraphs through hyperedge effective resistances, aiming to decompose hypergraphs into multiple node clusters with a small number of inter-cluster hyperedges. Building on the recent HyperEF framework, our approach offers three primary contributions. Specifically, first, by leveraging the expanded Krylov subspace exploiting both clique and star expansions of hyperedges, we can significantly improve the approximation accuracy of effective resistances. Second, we propose a resistance-based local clustering scheme for merging small isolated nodes into nearby clusters, yielding more balanced clusters with substantially improved conductance. Third, the proposed HyperEF 2.0 enables the integration of resistance-based hyperedge weighting and community detection into a multilevel hypergraph partitioning tool, achieving state-of-the-art performance. Extensive experiments on real-world VLSI benchmarks show that HyperEF 2.0 can more effectively coarsen hypergraphs without compromising their structural properties, while delivering much better solution quality (e.g. conductance) than the state-of-the-art hypergraph coarsening methods, such as HyperEF and HyperSF. Moreover, compared to leading hypergraph partitioners such as hMETIS, SpecPart, MedPart, and KaHyPar, our framework consistently achieves smaller cut sizes. In terms of runtime, HyperEF 2.0 attains up to a $4.5 \times$ speedup over the latest flow-based local clustering algorithm, HyperSF, demonstrating both superior efficiency and partitioning quality.
\end{abstract}


\section{Introduction}
The increasing complexity of modern networks necessitates efficient reduction techniques that maintain essential structural properties. Graph coarsening has become an indispensable technique for improving the scalability and effectiveness of algorithms in domains such as graph partitioning, embedding, and graph neural networks (GNNs). \cite{safro2015advanced, deng2019graphzoom, zhao2021towards, chen2022graph}.

Unlike simple graphs, hypergraphs naturally represent complex, higher-order relationships among entities \cite{HGreview}. However, most existing hypergraph coarsening methods rely on basic heuristics like vertex similarity or hyperedge similarity \cite{karypis1999multilevel, devine2006parallel, ccatalyurek2011patoh, Shaydulin_2019}. In hyperedge similarity-based methods, coarsening is achieved by merging large, similar hyperedges. While this simplifies implementation, it often distorts the original higher-order structure. On the other hand, vertex similarity-based methods attempt to group nodes based on their pairwise distances, typically computed from low-dimensional embeddings. While these methods are computationally efficient, they may only capture local relationships and overlook the global structural patterns that are crucial in hypergraphs. Consequently, these rudimentary metrics often fail to preserve the intricate connectivity and semantics inherent within the topological structure of the hypergraph.

Recent developments in spectral graph theory have facilitated sparsification and coarsening of simple graphs in nearly-linear time \cite{spielman2011spectral, feng2020grass, Lee:2017, zhuo:dac18, kapralov2022spectral, kapralov2021towards, zhang2020sf, lee2014multiway, spielmat1996spectral}. However, these methods do not directly extend to hypergraphs. Extending these ideas to hypergraphs often involves star or clique expansions \cite{hagen1992new}, which may lose multiway higher-order relationships and lead to lower performance. Alternatively, Soma et al. \cite{soma2018spectral} generalize spectral sparsification by sampling hyperedges with probabilities proportional to their weights relative to the minimum degree of any two vertices in the hyperedge. Another approach builds an explicit hypergraph Laplacian matrix \cite{zhou2006learning} and generalizes graph learning algorithms to the hypergraph domain. Chan et al. introduced a non-linear diffusion process to define the hypergraph Laplacian operator by measuring the flow distribution within each hyperedge \cite{chan2018spectral,chan2020generalizing}. Moreover, Cheeger’s inequality has been extended to hypergraphs using a diffusion-based nonlinear Laplacian operator \cite{chan2018spectral}. However, these theoretical advances do not readily yield efficient practical implementations. Recently, Ali et al.\ introduced HyperEF, a near-linear time spectral coarsening method that approximates hypergraph Laplacian eigenvalues via Krylov subspaces \cite{aghdaei2022hyperef}. While effective, it can underestimate important structural information and produce imbalanced clusters with isolated nodes.


In this work, we introduce HyperEF 2.0, a scalable spectral hypergraph coarsening approach that leverages effective-resistance clustering to produce substantially smaller yet structurally representative hypergraphs. HyperEF 2.0 effectively improves the solution quality of HyperEF by exploiting a mixed expansion scheme that adaptively selects Krylov subspace vectors derived from both star and clique expansions of hyperedges, to more accurately estimate hyperedge effective resistances. Additionally, we introduce a novel resistance-based local clustering strategy that merges isolated nodes into existing clusters by exploiting the structural properties of hypergraphs. This significantly enhanced hypergraph coarsening framework not only more accurately captures higher-order structures but also produces more balanced clusters, resulting in substantially improved outcomes across numerous real-world hypergraph partitioning tasks associated with modern VLSI design. The key contributions of this work are summarized as follows:
\paragraph{Enhanced Effective Resistance Estimation} We substantially improve the approximation of eigenvectors (in the Krylov subspace) by combining both clique and star expansions of hyperedges, capturing more nuanced structural information within the hypergraph.
\paragraph{Resistance-Based Local Clustering} We propose a local clustering technique that integrates small, isolated nodes into the most suitable clusters, leading to improved balance and enhanced conductance.
\paragraph{Integration into Partitioning Tools} By incorporating HyperEF 2.0 into resistance-based hyperedge weighting and community detection schemes,  we have developed a robust multilevel hypergraph partitioner that achieves superior cut sizes.
\paragraph{Extensive Empirical Validation} Our extensive experimental results, conducted with real-world VLSI designs, demonstrate that HyperEF 2.0 significantly enhances conductance and partition quality, while also achieving better runtimes compared to previous methods.

This paper is organized as follows: Section \ref{sec:background} provides an overview of key concepts in spectral hypergraph theory. Section \ref{sec:techniques} describes our proposed spectral coarsening method, including resistance-based clustering, local clustering, and its integration into hypergraph partitioning. Section \ref{sec:results} presents extensive experimental results across various benchmarks. Finally, Section \ref{sec:conclusion} concludes the paper, summarizing the key findings and their implications.

\section{Preliminaries and Background}\label{sec:background}
\subsubsection{Graph Laplacian matrix}
For an undirected graph $G = (\mathcal{V}, \mathcal{E}, z)$, with vertex set $\mathcal{V}$, edge set $\mathcal{E}$, and edge weights $z$, the adjacency matrix ${A}$ is defined as:
\begin{equation}\label{di_adjacency}
{A}(i,j)=\begin{cases}
z(i,j) & \text{if } (i,j)\in \mathcal{E},\\
0 & \text{otherwise}.
\end{cases}
\end{equation}
The Laplacian matrix of $G$ is then $L = D - A$, where ${D}$ is the diagonal degree matrix with ${D}(i,i)$ as the weighted degree of node $i$.
\subsubsection{Courant-Fischer Minimax Theorem}
For the Laplacian matrix $L \in \mathbb{R}^{|\mathcal{V}|\times|\mathcal{V}|}$, the $k$-th largest eigenvalue $\lambda_k(L)$ can be computed via:
\begin{equation}\label{eqn:minmax}
    \lambda_k(L) = \min_{dim(U)=k}\,{\max_{\substack{x \in U \\ x \neq 0}}{\frac{x^\top Lx}{x^\top x}}},
\end{equation}
where $U$ is a $k$-dimensional subspace of $\mathbb{R}^{\mathcal{V}}$. This can be utilized to determine the spectrum of the Laplacian matrix $L$.

\subsubsection{Hypergraph conductance}
A hypergraph $H = (V, E, w)$ comprises a vertex set $V$ and a set of hyperedges $E$ with weights $w$. The degree of a vertex $v$ is $d_v := \sum_{e \in E: v \in e} w(e)$. The volume of a node set $S \subseteq V$ is given by $vol(S) := \sum_{v \in S} d_v$. The conductance of $S$ is:
\begin{equation}
    \Phi(S) := \frac{cut(S, \hat{S})}{\min\{vol(S), vol(\hat{S})\}},
\end{equation}
where $cut(S, \hat{S})$ tracks how many hyperedges are split between $S$ and $\hat{S}$. The overall conductance of the hypergraph is:
\begin{equation}
    \Phi_H := \min\limits_{\emptyset \not\subseteq S \subseteq V} \Phi(S).
\end{equation}

\subsubsection{Cheeger's inequality}
Cheeger's inequality formalizes how closely a graph's conductance $\Phi_G$ is related to its spectral properties \cite{chung1997spectral}:
\begin{equation}\label{eqn:cheeger}
    \omega_2/2 \leq \Phi_G \leq \sqrt{2\omega_2},
\end{equation}
where $\omega_2$ is the second smallest eigenvalue of the normalized Laplacian matrix $\widetilde{L} = D^{-1/2} L D^{-1/2}$.

\subsubsection{Effective resistance distance}
Let $G = (\mathcal{V}, \mathcal{E}, z)$ be a connected, undirected simple graph with edge weights $z \in \mathbb{R}^\mathcal{E}_{\geq 0}$. Define the standard basis vector ${b_{p}} \in \mathbb{R}^{\mathcal{V}}$ to be zero everywhere except at node $p$, where it is one, and let ${b_{pq}}= b_p - b_q$. The effective resistance between nodes $p$ and $q$ is given by:
\begin{equation}\label{eq:eff_resist0}
    R_{eff}(p,q) = b_{pq}^\top L_G^{\dagger} b_{pq}=\sum\limits_{i= 2}^{|\mathcal{V}|} \frac{(u_i^\top b_{pq})^2}{\lambda_i} = \max\limits_{x \in \mathbb{R}^\mathcal{V}} \frac{(x^\top b_{pq})^2}{x^\top L_G x},
\end{equation}
where $L_G^{\dagger}$ is the Moore-Penrose pseudo-inverse of $L_G$, and $u_i$ denotes the unit-length, mutually-orthogonal eigenvectors of $L_G$ corresponding to Laplacian eigenvalues $\lambda_i$.

Intuitively, graph conductance measures how well-connected a subset of nodes is within a graph. A low conductance value indicates a tightly-knit cluster with few external edges. Similarly, in a graph modeled as a resistive network, lower effective resistance between two nodes indicates stronger connectivity due to the presence of multiple alternative paths.


\subsubsection{Hypergraph coarsening}
Multilevel coarsening methods aim to reduce the hypergraph's size by merging vertices based on rating functions. Alternatively, n-level approaches, such as KaSPar \cite{osipov2010n}, contract only one pair of vertices per level. KaHyPar \cite{kahypar} extends this technique to hypergraphs by applying the following rating function:
\begin{equation}
 \label{eq:rating_fuc}
    r(p,q) = \sum_{\{p,q\} \subseteq e , ~ e \in E}{\frac{w(e)}{|e| - 1}},
\end{equation}
where $w(e)$  is the hyperedge weight and $|e|$ is the hyperedge cardinality. This function prioritizes vertex pairs $(p,q)$ that are involved in hyperedges with relatively small cardinalities.

\subsubsection{Community Detection}
Although coarsening reduces a hypergraph’s size, essential structures can be lost during tie-breaking or when rating metrics are unclear. To address this, frameworks like KaHyPar use community detection before coarsening. In this approach, the hypergraph is divided into densely connected communities with sparse external links—often identified by maximizing a modularity-based objective (e.g., via the Louvain algorithm). Coarsening is then applied within each community to preserve structural details more effectively.

\subsubsection{Partitioning objectives}
Hypergraph partitioning generalizes graph partitioning by dividing the vertex set into $k$ parts subject to two constraints:
\begin{itemize}
\item Each part maintains nearly the same total vertex weight, satisfying $(\frac{1}{k}- \epsilon)W \leq \sum_{v \in V_i}{w_v} \leq (\frac{1}{k}+\epsilon)W$ for each $V_i$.
\item The overall cut size, $cutsize_H{(S)} = \sum_{\{e|e \not\subseteq V_i \text{ for any }i\}}{w_e}$ is minimized.
\end{itemize}
Here, $W = \sum_{v \in V}{w_v}$ is the total vertex weight, $\epsilon$  is a small imbalance tolerance, and $w_e$ denotes the weight of hyperedge $e$. The end goal is an $\epsilon$-balanced partition $S = \{V_0, V_1, \dots, V_{k-1}\}$ that achieves minimal cut size.

\section{Spectral Coarsening via Enhanced Resistance Estimation and Local Clustering}\label{sec:techniques}

We overcome the limitations of previous hypergraph coarsening methods, which rely on simplistic heuristics, by introducing a theoretically grounded and effective spectral coarsening framework. Building on recent advances in spectral hypergraph clustering \cite{aghdaei2022hyperef}, our approach, HyperEF 2.0, first utilizes a mixed expansion scheme of hyperedges to improve the accuracy of resistance approximation. This is followed by a resistance-based local clustering technique to enhance both clustering quality and balance. Finally, we integrate this coarsening scheme into a multilevel hypergraph partitioning framework to achieve a substantially improved solution.
\subsection{Resistance-Based Hypergraph Clustering}
We utilize an effective-resistance measure to iteratively coarsen hyperedges by contracting those with low effective resistance. While effective resistance has been applied in simple graphs to identify critical edges and evaluate overall connectivity, it has received less attention in the context of hypergraphs.

Existing coarsening algorithms typically rely on local rating functions—such as those based solely on hyperedge size or weight—to merge vertices at each hierarchy level. However, these local metrics can overlook globally significant structures. For example, when a hyperedge functions as a "bridge," size-based methods may mistakenly merge its nodes, potentially collapsing the hypergraph’s structure. In contrast, a high effective resistance for bridging hyperedges prevents such contractions, thus preserving global connectivity.
\subsubsection{Hypergraph Clustering by Effective Resistances} 
The proposed spectral hypergraph coarsening strategy, similar to HyperEF \cite{aghdaei2022hyperef}, groups nodes within each hyperedge if they exhibit a low effective-resistance diameter (see Fig. \ref{fig:EF_overview}). This approach significantly reduces the hypergraph's size while preserving key structural characteristics of the original. The core feature of HyperEF 2.0 is an efficient algorithm for estimating hyperedge effective resistances, which adapts the optimization-based formulation from Eq. (\ref{eq:eff_resist0}) to hypergraphs.  Concretely, we determine the effective resistance of a hyperedge by solving for the optimal vector $\chi^*$ in the following maximization: \begin{figure}
    \centering
    \includegraphics [width = 0.6\linewidth]{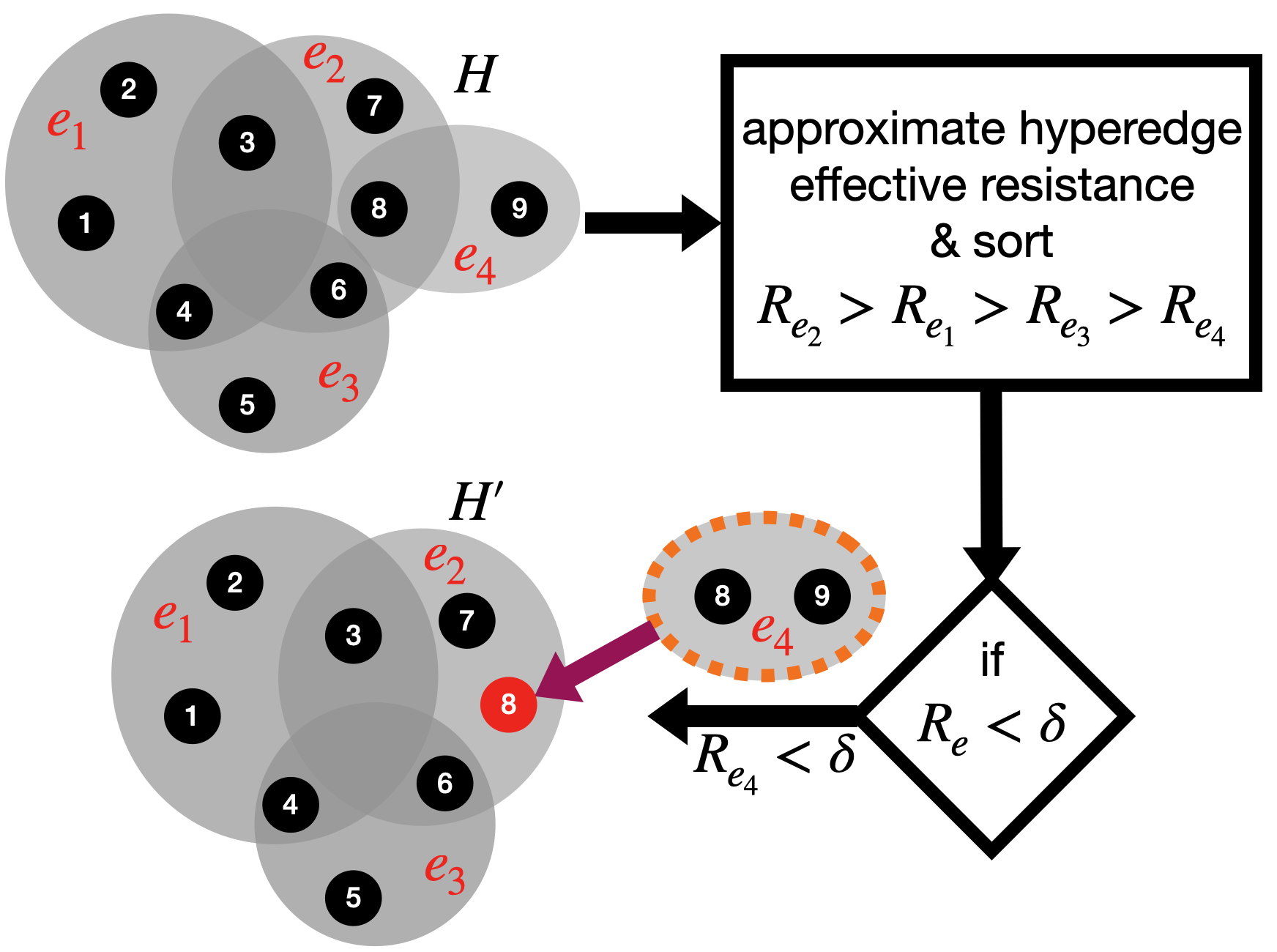}
    \caption{Overview of the HyperEF method \cite{aghdaei2022hyperef}.}
    \label{fig:EF_overview}
\end{figure}

\begin{equation} \label{eq:eff_resist_hypergraph}
\begin{split}
  R_e(\chi^*) =\max\limits_{\chi \in \mathbb{R}^\mathcal{V}} \frac{(\chi^{\top} b_{pq})^2}{Q_H(\chi)},  \quad p,q\in e
\end{split}
\end{equation}
where the original quadratic form $x^\top L_G x$ from Eq. (\ref{eq:eff_resist0}) is replaced by the nonlinear quadratic form $Q_H(\chi)$ \cite{chan2018spectral}: \begin{equation}\label{eq:non-linerQ}
    Q_H(\chi) := \sum\limits_{e\in E} w_e \max\limits_{u, v \in e} (\chi_u - \chi_v)^2.
\end{equation}

As illustrated in Fig. \ref{fig:EF_overview}, this method produces a significantly reduced hypergraph $H' = (V', E', w')$ from the original $H = (V, E, w)$ by leveraging hyperedge effective resistances, thereby reducing the number of vertices, edges, and overall weights ($|V'| < |V|$, $|E'| < |E|$ and $|w'|<|w|$).

\subsubsection{Low-Resistance-Diameter Hypergraph Decomposition}  
Let $G = (\mathcal{V}, \mathcal{E}, z)$ be a weighted, undirected graph with positive edge weights $z$ and some sufficiently large $\gamma > 1$. The effective resistance diameter is defined as $\max\limits_{u,v \in \mathcal{V}} R_{eff}(u,v)$. Recent work shows that one can create multiple node clusters $G[\mathcal{V}_i]$ each with low effective-resistance diameters by discarding only a small fraction of the edges \cite{alev2018graph}:
\begin{equation}\label{them:res}
    \max\limits_{u,v \in \mathcal{V}_i} R_{eff_{G[\mathcal{V}_i]}}(u,v) \lesssim \gamma^3 \frac{|\mathcal{V}|}{z(\mathcal{E})}.
\end{equation}
Moreover, let $\Phi_G$  denote the conductance of $G$. By Cheeger's inequality, one can bound the effective-resistance diameter in terms of the graph’s conductance \cite{alev2018graph}:
\begin{equation}\label{them:res1}
     \max\limits_{u,v \in \mathcal{V}}R_{eff}(u,v) \lesssim \frac{1}{\Phi_G^2}.
\end{equation}

These results extend naturally to hypergraphs \cite{chan2018spectral,chan2020generalizing}. Inequality (\ref{them:res}) indicates that a hypergraph can be decomposed into multiple hyperedge clusters of small effective-resistance diameter by removing only a few inter-cluster hyperedges. Meanwhile, (\ref{them:res1}) implies that contracting hyperedges (node clusters) with low effective-resistance diameter has minimal impact on the hypergraph’s overall conductance.

\subsubsection{\textbf{Enhanced Estimation of Hyperedge Effective Resistances}}
To efficiently approximate the optimal vector $\chi^*$ in Eq. (\ref{eq:eff_resist_hypergraph}), we restrict the search space to an eigensubspace spanned by a select set of orthogonal Laplacian eigenvectors derived from simplified graph representations of the hypergraph. Let $G_b = (\mathcal{V}_b, \mathcal{E}_b, z_b)$ denote the bipartite graph representation of the hypergraph $H = (V, E, w)$, where $|\mathcal{V}_b| = |V| + |E|$ and $|\mathcal{E}_b| = \sum_{e \in E} |e|$, with edge weights defined as $z_b(e, p) = \frac{w(e)}{d(e)}$. We also introduce a complementary representation: the clique expansion graph $G_c = (\mathcal{V}_c, \mathcal{E}_c, z_c)$, where $|\mathcal{V}_c| = |V|$ and edges exist between all node pairs within each hyperedge. The weight function $z_c(u, v) = \sum_{e \in E; u,v \in e} \frac{w(e)}{\binom{|e|}{2}}$ ensures that the weight of each hyperedge is evenly distributed among all node pairs it contains. These dual representations—star and clique expansions—enable a more comprehensive capture of structural details than either method alone.

Building on the foundation of HyperEF \cite{aghdaei2022hyperef}, we exploit a Krylov subspace approach to approximate the eigenvectors. For a nonsingular matrix $A \in \mathbb{R}^{n \times n}$ and a non-zero vector $x \in \mathbb{R}^n$, the order-$(\rho+1)$ Krylov subspace is defined as: \begin{equation}\label{eq:krylov} \kappa_{\rho}(A, x) := \text{span}(x, Ax, A^2x, ..., A^{\rho}x), \end{equation} where $A$ is the normalized adjacency matrix of either $G_b$ or $G_c$, and $x$ is a random vector. In contrast to HyperEF, which relies solely on star expansion, our method integrates Krylov subspace vectors from both star and clique expansions to construct a unified pool of mutually orthogonal vectors. While incorporating the clique expansion introduces some overhead, empirical results show that it more effectively preserves high-order hyperedge structure and isolates critical hyperedges by amplifying their resistance scores.

Note that these Krylov subspace vectors are computed once per hypergraph using sparse matrix-vector operations, yielding a pool of $2\rho$ orthogonal embeddings. From this pool, only the vectors that maximize Eq. (\ref{eq:eff_resist_hypergraph}) will be selected for hyperedge resistance estimation. By removing the vector entries corresponding to star nodes, we form candidate vectors $\chi^{(1)}, ..., \chi^{(2\rho)}$, each of which allows embedding the hypergraph nodes into a $2\rho$-dimensional space. For each hyperedge $e$, we compute its resistance ratio as:
\begin{equation}\label{eq:ratios} r_e(\chi^{(i)}) = \frac{(\chi^{(i)\top} b_{pq})^2}{Q_H(\chi^{(i)})}, \quad p,q \in e, \end{equation}
where $p$ and $q$ are the most distant nodes in the embedding space. Finally, we estimate the effective resistance of $e$ as the maximum: \begin{equation}\label{eq:R} R_e = \max_{i=1,...,2\rho} r_e^{(i)}, \quad e \in E. \end{equation} This refinement scheme enables more effective capture of structurally important hyperedges while maintaining computational efficiency. Algorithm \ref{alg:effR} outlines the proposed method for estimating hyperedge effective resistance.
\begin{algorithm}
\small { \caption{Effective resistance estimation}\label{alg:effR}}
\textbf{Input:} Hypergraph $H = (V,E, w)$.\\
\textbf{Output:} {Hyperedge's effective resistance vector $R$}.\\
  \algsetup{indent=1em, linenosize=\small} \algsetup{indent=1em}
    \begin{algorithmic}[1]
    \STATE Construct a bipartite graph $G_b$ corresponding to $H$.
    \STATE Construct a clique expansion $G_c$ corresponding to $H$.
    \STATE Construct a Krylov subspace from the combined vector pool of both expansions to capture the details.
    \STATE Use the Gram–Schmidt method to obtain the orthogonal vectors.
    \STATE For each hyperedge, calculate its $2\rho$ resistance ratios  using (\ref{eq:ratios}).
    \STATE Sort hyperedges based on their resistance ratios. 
    \STATE Calculate all hyperedge effective resistances $R$ based on (\ref{eq:R}).
     \STATE Return  $R$. 
    \end{algorithmic}
\end{algorithm}

To ensure high efficiency, the proposed spectral hypergraph coarsening framework utilizes a linear-time local spectral embedding technique that applies low-pass filtering (smoothing) to random graph signals, adapted for hypergraphs \cite{cohen2017almost, cohen:focs18, soma2018spectral, chan2018spectral}. The coarsening algorithms are also designed with parallel-friendly sparse operations, making them well-suited for acceleration on modern hardware architectures \cite{bell2008efficient, greathouse2014efficient, fowers2014high, buono2016optimizing, hong2018efficient, zhang2020sparch}.

\subsubsection{Multilevel Effective Resistance Clustering}
To strengthen structural fidelity during coarsening, we adopt a multilevel clustering approach inspired by HyperEF, in which hyperedges with low effective resistance ($R_e < \delta$) are iteratively contracted. At each coarsening level, nodes within a cluster are merged into a new \textit{supernode}. We then assign a weight to each supernode equal to the effective resistance of the hyperedge that formed it at the previous level. This weighting scheme is crucial for propagating essential structural information throughout the entire hierarchy.

Let $H^{(l)} = (V^{(l)}, E^{(l)}, w^{(l)})$ denote the hypergraph at level $l$. The term $\eta(v)$ represents the weight of the nodes $v \in e$ corresponding to a contracted hyperedge from the previous level, initially set to all zeros for the original hypergraph. As a result, the effective resistance of a hyperedge at a coarser level is determined by both the resistance calculated at the current level, $R_e^{(l)}$,  and the resistance data accumulated from all previous levels:
\begin{equation}\label{eq:R_W_update}
     R_e^{(l)} \gets \sum_{v \in e} \eta(v)+R_e^{(l)},
\end{equation}
This formulation ensures that historical resistance data informs current decisions, allowing the algorithm to preserve global structural information across all coarsening levels.

\subsection{Resistance-based Local Clustering}
To more effectively address isolated nodes resulting from the resistance-based edge contraction phase, we introduce an efficient and effective local clustering approach. This method significantly improves coarsening results and leads to more balanced node clusters. Compared to the recent HyperSF \cite{aghdaei2021hypersf} algorithm, which relies on a local flow-based method \cite{veldt2020minimizing}, the proposed resistance-based local clustering approach has demonstrated more promising results, offering better solution quality and reduced runtime in hypergraph coarsening and partitioning tasks for real-world VLSI designs.

\begin{figure*}[!t]
\centering
\includegraphics[width=0.899\textwidth]{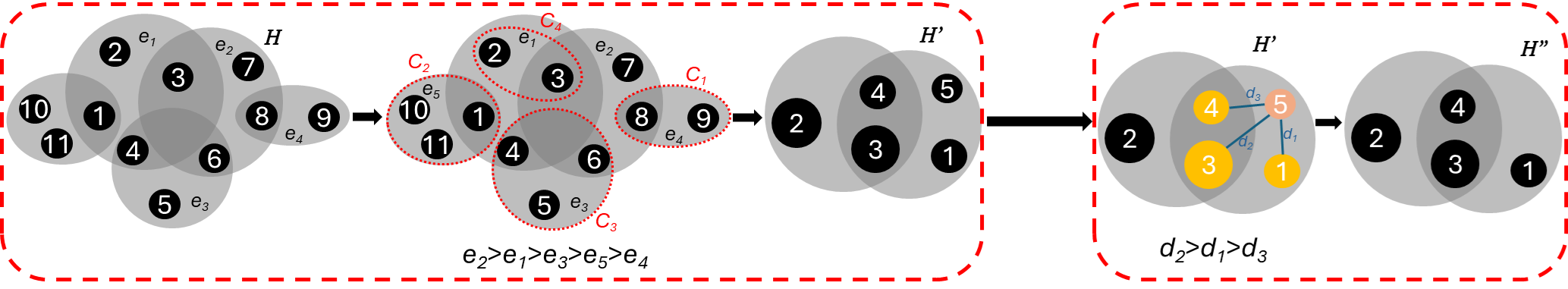}
\caption{HyperEF 2.0 includes two key steps: (1) resistance-based hyperedge contraction, and (2) resistance-based local clustering.}
\label{fig:hyperef}
\end{figure*}

\subsubsection{Overview of Coarsening Refinement}
Fig. \ref{fig:hyperef} illustrates our proposed refinement strategy. We use a local clustering algorithm to refine heavily imbalanced node clusters identified at each coarsening level, which typically have notably smaller resistance diameters compared to other clusters. Utilizing this algorithm, HyperEF 2.0 repeatedly detects sets of adjacent node clusters for each selected node, aiming to find the minimum effective resistance distance in Eq. (\ref{eq:ratios}) between small “isolated” clusters and their neighbors. The following steps outline the main procedure:
\textbf{(Step 1)} Identify isolated (\textit{super}) nodes. 
\textbf{(Step 2)} Evaluate neighboring clusters for each isolated node.
\textbf{(Step 3)} Compute effective-resistance distance between the isolated node (or cluster) and each neighbor.
\textbf{(Step 4)} Find the best neighbor with the minimum resistance distance to the isolated node.
\textbf{(Step 5)} Merge the isolated cluster with its best neighbor to construct a more balanced coarsening scheme. The effective-resistance diameter of the newly formed cluster is then updated using Eq.~(\ref{eq:R_W_update}) for propagating structural information to the next coarsening level.
One key advantage of this method is its ability to minimize information loss: by merging isolated nodes as soon as they are detected, we preserve crucial connectivity details that could otherwise be lost during the coarsening phase.

\subsubsection{Local Clustering Algorithm} Our proposed algorithm is strongly local, expanding the hypergraph around seed nodes $\mathcal{C}$. This design offers two key benefits for coarsening: (1) restricting node aggregation to the local neighborhoods of seed nodes preserves the hypergraph’s global structure, and (2) limiting the clustering to smaller subproblems greatly enhances runtime efficiency.

First, we apply Algorithm~\ref{alg:effR} to the hypergraph $H=(V, E, w)$ to estimate hyperedge effective resistances. This is followed by multilevel effective resistance clustering, which produces a coarsened hypergraph $H' = (V', E', w')$ and identifies unclustered (isolated) nodes, denoted by $\mathcal{C}$. These isolated nodes are simply those that remain ungrouped after the multilevel coarsening step. HyperEF 2.0 then constructs a sub-hypergraph $H'_L$ around the seed nodes $\mathcal{C}$ to find the best neighbor with the smallest distance using Krylov subspace-based vectors from Eq. (\ref{eq:krylov}) along with Eqs. (\ref{eq:ratios}) and (\ref{eq:R}).

To compute the distance between each isolated node $\mathcal{C}$ and its neighboring clusters, we reuse the vector $\chi$ generated from Eq.~(\ref{eq:krylov}) to evaluate the numerator of Eq.~(\ref{eq:ratios}). In this context, $p$ refers to the isolated node, and $q$ corresponds to a cluster (supernode). To represent the cluster as a single node in the embedding space, we define its vector as the average of the $\chi$-vectors of all nodes within that cluster. This average can be precomputed once per cluster for efficiency. We then calculate the distance $d$ between the isolated node and each cluster, sort these distances in descending order (e.g., $d_2 > d_1 > d_3$, as shown in Fig.~\ref{fig:hyperef}), and assign the isolated node to the cluster with the smallest distance.

\begin{figure}[bp]
    \centering
         \includegraphics [  width=0.8\linewidth]{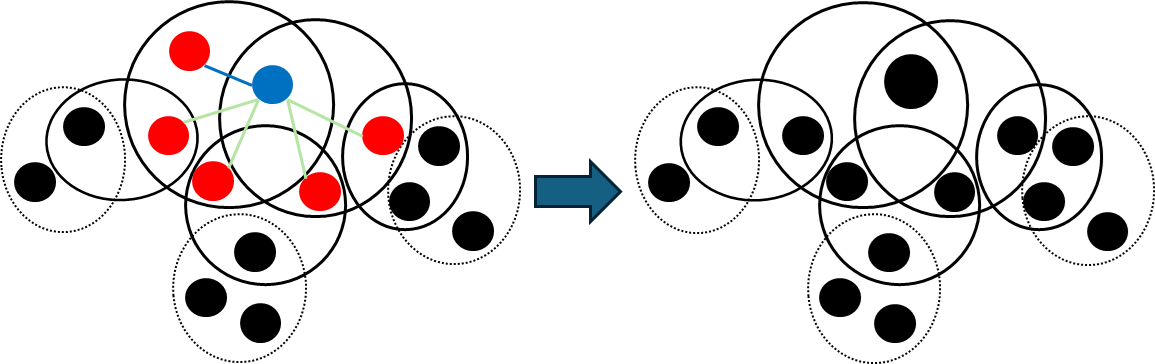}
    \caption{Constructing sub-hypergraph through local clustering}
    \label{fig:hypersf}
\end{figure}

Starting from the current hypergraph $H' = (V', E', w')$  and an isolated seed node $\mathcal{C} \in V'$,  we collect all incident hyperedges $E'(\mathcal{C}) = \{ e' \in E' ~|~ \mathcal{C}\in e' \}$ and treat the co-occurring vertices as the seed’s neighborhood:
\begin{equation}\label{eq:neighbors}
    \kappa (\mathcal{C}) = \cup_{e' \in E'(\mathcal{C})}(e' ~\backslash \{\mathcal{C}\}).
\end{equation}
This yields the local vertex set $V'_L = \{\mathcal{C}\} \cup \kappa (\mathcal{C})$ and the corresponding edge set $E'_L = \{ e' \in E' ~|~ e' \subseteq V'_L\}$; together they define the compact sub‑hypergraph $H'_L = (V'_L, E'_L, w'_L)$. As shown in Fig.~\ref{fig:hypersf}, HyperEF 2.0 confines its search for the best neighboring cluster to $H'_L$, ensuring that distance calculations remain focused on the immediate vicinity of the seed while avoiding a full-hypergraph scan.

Algorithm \ref{alg:HyperEF} outlines the details of the proposed resistance-based local clustering technique, HyperEF 2.0. It takes the original hypergraph $H$ as input and produces a set of strongly connected vertices as output.

\begin{algorithm}
\small { \caption{Resistance-Based Hypergraph Local Clustering}\label{alg:HyperEF}}
\textbf{Input:} Hypergraph $H = (V,E, w)$.\\
\textbf{Output:} {A coarsened hypergraph $H' = (V', E', w')$ and vertex clusters $C$.}\\
  \algsetup{indent=1em, linenosize=\small} \algsetup{indent=1em}
    \begin{algorithmic}[1]
    \STATE Initialize $H'$ $\leftarrow$ $H$
    \FOR{each coarsening level $L$}
    \STATE Call Algorithm \ref{alg:effR} to compute the effective‑resistance vector $R$ for $H'$.
    \STATE Update the effective resistance vector $R$ using Eq. (\ref{eq:R_W_update})
    \STATE Starting with the hyperedges that have the lowest effective resistances, contract the hyperedge if $R_e< \delta$.
    \STATE Rebuild the corresponding coarsened hypergraph $H'$.
    \STATE Identify isolated nodes $\mathcal{C} \subseteq V$ and assign them as seed nodes $S \gets \mathcal{C}$;
    \STATE Construct the local sub-hypergraph $H'_L$ around seed nodes.
    \STATE Use computed vectors to find the best neighboring clusters $\kappa(S)$ for the seeds.
    \STATE Merge each seed node with its best neighboring cluster and update effective resistances accordingly. 
    \ENDFOR
    \STATE Return $H'$, $C$.
    \end{algorithmic}
\end{algorithm}

\subsection{Algorithm Complexity}
In HyperEF 2.0, constructing the Krylov subspace for the bipartite graph $G_b$ (star expansion) takes $\mathcal{O}(|\mathcal{E}_b|)$ time, where $|\mathcal{E}_b| = \sum_{e \in E} |e|$. For the clique graph $G_c$, the worst-case construction time is $\mathcal{O}(\sum_{e \in E} k_e^2)$, where $k_e = |e|$. When hyperedge sizes are bounded ($\Delta \ll n$), this reduces to $\mathcal{O}(p\Delta)$ with $p = \sum_{e \in E} k_e$ and $\Delta = \max_{e \in E} k_e$. Hyperedge resistance estimation and clustering each require $\mathcal{O}(\rho |E|)$ time, while node weight computation and resistance-based refinement both take $\mathcal{O}(|E|)$ time. Overall, the runtime is: $\mathcal{O}(\rho|E|+|\mathcal{E}_b| + \sum\limits_{e \in E} k_e^2)$

\subsection{HyperEF 2.0 for Multilevel Hypergraph Partitioning}
To incorporate HyperEF 2.0 into hypergraph partitioning workflows, we replace conventional heuristic coarsening with our resistance-based multilevel spectral coarsening approach and use HyperEF 2.0 clustering for community detection, thereby enhancing partitioning quality.

\subsubsection{Multilevel spectral  coarsening}
Most multilevel schemes employ a rating function at each level. We propose a new function derived from effective resistance, selecting vertex pairs ($p,q$) in hyperedges that feature heavy nets and low effective resistance:
\begin{equation}\label{eq:ratingF}
    r(p,q) = \sum_{\{p,q\} \subseteq e,~ e \in E}{\frac{w(e)}{R_e - 1}},
\end{equation}
where $w(e)$  denotes the weight of hyperedge $e$, and $R_e$ is the hyperedge’s effective resistance (Eq.~(\ref{eq:R})). Under this scheme, each coarse-level vertex is assigned a weight equal to the effective-resistance diameter of its corresponding cluster in the finer level. Consequently, hyperedge resistances at coarser levels reflect both newly evaluated values and data passed on from all previous levels, preserving crucial structural information.
\begin{figure}[htbp]
    \centering
         \includegraphics [  width=0.8\linewidth]{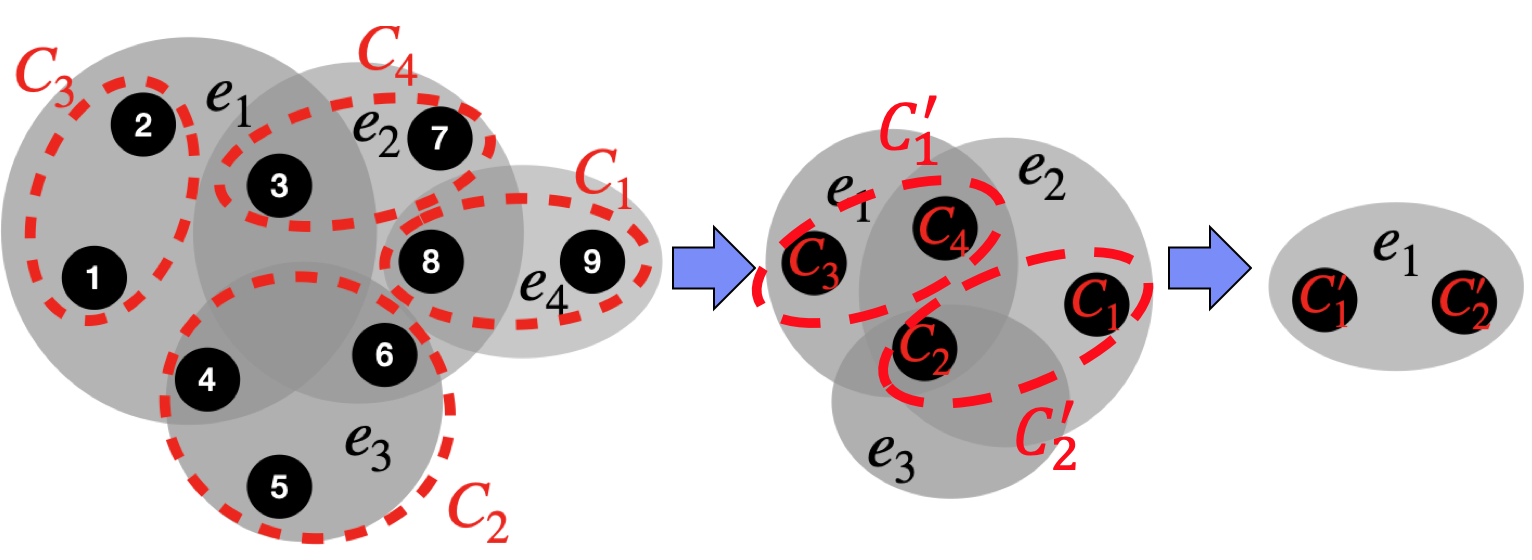}
    \caption{Multilevel spectral coarsening \cite{aghdaei2022hyperef}. \protect\label{fig:multilevel}}
\end{figure}

\subsubsection{Community Detection with HyperEF 2.0} HyperEF 2.0 performs clustering as a refinement step to improve partition quality. It begins by initializing isolated clusters, then merges nearby clusters with low resistance diameters within their shared neighborhoods. This process results in clusters with reduced average conductance, thereby better preserving the hypergraph’s spectral properties. In practice, we treat community detection as a hypergraph clustering problem—first identifying node clusters, then applying coarsening within each community to achieve both fine-grained accuracy and global consistency.

In Algorithm~\ref{alg:Shypar}, we outline the detailed steps of our hypergraph partitioner, which employs resistance-based spectral coarsening and clustering to enhance partitioning performance.

\begin{algorithm}
\small { \caption{Hypergraph Partitioning  with HyperEF 2.0}\label{alg:Shypar}}
\textbf{Input:} Hypergraph $H = (V,E,w)$ \\
\textbf{Output:} {Partitioned Hypergraph};\\
    \algsetup{indent=1em, linenosize=\small} \algsetup{indent=1em}
    \begin{algorithmic}[1]
    \STATE Setup: Hyperedge Effective resistance computation using algorithm {\ref{alg:effR}}
    \STATE Setup: Community detection using HyperEF 2.0 clustering in algorithm {\ref{alg:HyperEF}}
    \STATE Multilevel spectral coarsening using Eq. ({\ref{eq:ratingF}})
    \STATE Initial partitioning based on KaHyPar algorithm.
    \STATE Solution refinement based on KaHyPar algorithm.
    \STATE Return the partitioned hypergraph 
    \end{algorithmic}
\end{algorithm}

\section{Experimental Validation}\label{sec:results}
Hypergraph models are widely used in various VLSI design tasks. For example, chip placement, a critical stage in modern VLSI physical design, directly influences key design quality metrics such as timing closure, die usage, and routability.
In this work, we provide a comprehensive evaluation of our hypergraph clustering framework for VLSI CAD applications, focusing on both solution quality and runtime efficiency. We test our approach on a broad range of public-domain datasets. The implementation of the proposed algorithm is publicly available at \url{https://github.com/Feng-Research/HyperEF2.0}.

Specifically, we apply our clustering method and multilevel hypergraph partitioning tools to well-known VLSI design benchmarks, including the ISPD98 suite \cite{alpert1998ispd98} and the Titan23 dataset \cite{murray2013titan}. Details of these benchmarks are provided in Tables~\ref{tab:table1} and~\ref{tab:table_Titan23}. Additionally, we apply HyperEF 2.0’s coarsening and clustering techniques to the hypergraph partitioning problem and compare the partitioning results with those from leading hypergraph partitioners—hMETIS \cite{karypis1999multilevel}, SpecPart \cite{bustany2022specpart}, KaHyPar \cite{kahypar}, and MedPart \cite{liang2024medpart}—using the aforementioned benchmark suites.

\subsection{Spectral Coarsening with HyperEF 2.0} In this section, we compare our resistance-based hypergraph clustering approach against the widely used hMETIS \cite{karypis1999multilevel}, using real-world VLSI design benchmarks \cite{alpert1998ispd98}. All experiments were conducted on a server equipped with an Intel(R) Xeon(R) Gold 6244 processor and 1546 GB of memory.

\subsubsection{HyperEF 2.0 vs.\ hMETIS for Hypergraph Coarsening} We evaluate both solution quality and runtime efficiency. To measure solution quality, we use the average conductance of the node clusters: \begin{equation}
    \Phi_{\text{avg}} = \frac{1}{\left| S\right|}\sum\limits_{i=1}^{\left| S\right|} \Phi(S_i)
\end{equation}
where $\Phi(S_i)$ is the conductance of cluster $S_i$. Tables \ref{tab:hmetis} and \ref{tab:hmetis1} further compare the average conductance $\Phi_{\text{avg}}$ of node clusters generated by HyperEF 2.0 and hMETIS, under the same node reduction ratios (NRs). With an NR of 60\% (a $2.5\times$ node reduction), HyperEF 2.0 outperforms hMETIS by up to 15\% in $\Phi_{\text{avg}}$; at an NR of 80\%, the improvement is up to 12\%. HyperEF 2.0 also delivers speedups of up to $16\times$ over hMETIS.

We note that, although a direct comparison between HyperEF 2.0 and the original HyperEF algorithm \cite{aghdaei2022hyperef} is not possible due to the different number of clusters produced, we can still observe a substantial improvement in clustering performance when comparing them with hMETIS \cite{hmetis}. Specifically, compared to the original HyperEF algorithm, HyperEF 2.0 achieves nearly twice the improvement in average conductance over hMETIS, highlighting the effectiveness of our enhanced coarsening strategy.

\begin{table}[ht]
\caption{HyperEF 2.0 vs hMETIS conductance (NR=60\%)} \label{tab:hmetis}
\centering
\resizebox{.4\textwidth}{!}{
\footnotesize
\begin{tabular}{|c|c|c|c|c|}
\hline
Benchmark &
\begin{tabular}{@{}c@{}} $\Phi_{\text{avg}}$ \\
HyperEF 2.0
\end{tabular}      & \begin{tabular}{@{}c@{}} $\Phi_{\text{avg}}$ \\
hMETIS
\end{tabular}       & \begin{tabular}{@{}c@{}} t (second) \\
HyperEF 2.0
\end{tabular}   & \begin{tabular}{@{}c@{}} t (second) \\
hMETIS
\end{tabular}             \\ \hline
IBM01  & {\textbf{0.67}} & 0.75 & 9.1 & 10 $(\times 1)$ \\ \hline
IBM02  & {\textbf{0.67}} & 0.78 & 10 & 27 $(\times 3)$ \\ \hline
IBM03  & {\textbf{0.68}} & 0.76 & 10.3 & 33 $(\times 3)$ \\ \hline
IBM04  & {\textbf{0.68}} & 0.76 & 10.7 & 38 $(\times 4)$ \\ \hline
IBM05  & {\textbf{0.62}} & 0.73 & 11.1 & 42 $(\times 4)$ \\ \hline
IBM06  & {\textbf{0.70}} & 0.77 & 11.1 & 50 $(\times 4)$  \\ \hline
IBM07  & {\textbf{0.69}} & 0.77 & 11.6 & 78 $(\times 7)$ \\ \hline
IBM08  & {\textbf{0.69}} & 0.78 & 12.6 & 81 $(\times 6)$ \\ \hline
IBM09  & {\textbf{0.69}} & 0.78 & 12.5 & 85 $(\times 7)$ \\ \hline
IBM10  & {\textbf{0.67}} & 0.77 & 14.6 & 121 $(\times 8)$ \\ \hline
IBM11  & {\textbf{0.68}} & 0.77 & 14.1 & 120 $(\times 8)$  \\ \hline
IBM12  & {\textbf{0.69}} & 0.79 & 14.6 & 137 $(\times 9)$ \\ \hline
IBM13  & {\textbf{0.70}} & 0.78 & 16 & 142 $(\times 9)$ \\ \hline
IBM14  & {\textbf{0.68}} & 0.77 & 22.1 & 315 $(\times 14)$ \\ \hline
IBM15  & {\textbf{0.71}} & 0.79 & 28.2 & 349 $(\times 12)$  \\ \hline
IBM16  & {\textbf{0.69}} & 0.78 & 30.3 & 408 $(\times 13)$  \\ \hline
IBM17  & {\textbf{0.70}} & 0.80 & 31.5 & 505 $(\times 16)$  \\ \hline
IBM18  & {\textbf{0.68}} & 0.78 & 33.3 & 425 $(\times 13)$  \\ \hline
\end{tabular}
}
\end{table}

\begin{table}[ht]
\caption{HyperEF 2.0 vs hMETIS conductance (NR=80\%)} \label{tab:hmetis1}
\centering
\resizebox{.4\textwidth}{!}{
\footnotesize
\begin{tabular}{|c|c|c|c|c|}
\hline
Benchmark &
\begin{tabular}{@{}c@{}} $\Phi_{\text{avg}}$ \\
HyperEF 2.0
\end{tabular}      & \begin{tabular}{@{}c@{}} $\Phi_{\text{avg}}$ \\
hMETIS
\end{tabular}       & \begin{tabular}{@{}c@{}} t (second) \\
HyperEF 2.0
\end{tabular}   & \begin{tabular}{@{}c@{}} t (second) \\
hMETIS
\end{tabular}  \\ \hline
IBM01  & {\textbf{0.53}} & 0.59 & 10 & 10 $(\times 1)$ \\ \hline
IBM02  & {\textbf{0.57}} & 0.65 & 11.9 & 26 $(\times 2)$ \\ \hline
IBM03  & {\textbf{0.57}} & 0.61 & 12.3 & 32 $(\times 3)$ \\ \hline
IBM04  & {\textbf{0.57}} & 0.61 & 12.8 & 37 $(\times 3)$ \\ \hline
IBM05  & {\textbf{0.53}} & 0.59 & 13.2 & 41 $(\times 3)$ \\ \hline
IBM06  & {\textbf{0.58}} & 0.63 & 13.8 & 48 $(\times 3)$  \\ \hline
IBM07  & {\textbf{0.56}} & 0.64 & 14.2 & 76 $(\times 5)$ \\ \hline
IBM08  & {\textbf{0.57}} & 0.64 & 15.2 & 78 $(\times 5)$ \\ \hline
IBM09  & {\textbf{0.57}} & 0.63 & 15.2 & 82 $(\times 5)$ \\ \hline
IBM10  & {\textbf{0.55}} & 0.63 & 17.5 & 117 $(\times 7)$ \\ \hline
IBM11  & {\textbf{0.56}} & 0.63 & 17.2 & 116 $(\times 7)$   \\ \hline
IBM12  & {\textbf{0.58}} & 0.66 & 18 & 132 $(\times 7)$  \\ \hline
IBM13  & {\textbf{0.59}} & 0.64 & 20 & 138 $(\times 7)$  \\ \hline
IBM14  & {\textbf{0.56}} & 0.64 & 25.1 & 304 $(\times 12)$ \\ \hline
IBM15  & {\textbf{0.59}} & 0.65 & 34 & 337 $(\times 10)$  \\ \hline
IBM16  & {\textbf{0.57}} & 0.64 & 36.1 & 394 $(\times 11)$  \\ \hline
IBM17  & {\textbf{0.59}} & 0.67 & 37.8 & 488 $(\times 13)$  \\ \hline
IBM18  & {\textbf{0.56}} & 0.63 & 40.1 & 411 $(\times 10)$  \\ \hline
\end{tabular}
}
\end{table}

\subsection{Hypergraph Partitioning with Spectral Coarsening} 
\subsubsection{Experimental Setup}
To implement the hypergraph partitioner based on our proposed method, we developed hypergraph partitioning tools using the open-source multilevel hypergraph partitioner KaHyPar, integrating our spectral hypergraph coarsening and clustering techniques. Specifically, we replaced KaHyPar’s heuristic coarsening scheme with our spectral coarsening algorithm to form a hierarchy of coarser hypergraphs, alongside a new rating function. In addition, we substituted the Louvain-based community detection method in KaHyPar with our local clustering approach.

\subsubsection{Partitioning performance}
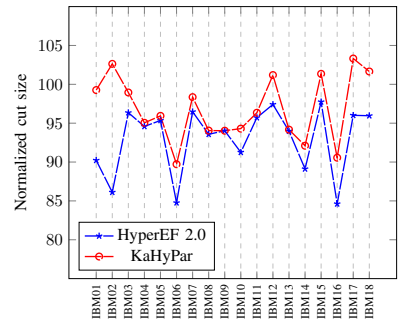
\begin{figure}[bp]
    \centering
    \resizebox{0.6\linewidth}{!}{
\begin{tikzpicture}
\begin{axis}[
    ylabel={Normalized cut size},
    symbolic x coords={IBM01, IBM02, IBM03, IBM04, IBM05, IBM06, IBM07, IBM08, IBM09, IBM10, IBM11, IBM12, IBM13, IBM14, IBM15, IBM16, IBM17, IBM18},
    xtick=data,
    x tick label style={rotate=90, anchor=east, font=\scriptsize}, 
    ytick={80,85,...,105},
    ymin=75,
    ymax=110,
    legend pos = south west,
    ymajorgrids=false,
    xmajorgrids=true,
    grid style=dashed,
]

\addplot[
    color=blue,
    mark=star,
    style=thick,
    dash pattern=on 10pt off 1pt
    ]
    coordinates {
    (IBM01,90.22)(IBM02,86.11)(IBM03,96.35)(IBM04,94.58)(IBM05,95.35)(IBM06,84.76)(IBM07,96.47)(IBM08,93.57)(IBM09,94.03)(IBM10,91.25)(IBM11,95.71)(IBM12,97.41)(IBM13,93.98)(IBM14,89.12)(IBM15,97.74)(IBM16,84.62)(IBM17,96.01)(IBM18,95.96)
};
\addlegendentry{HyperEF 2.0}

\addplot[
    color=red,
    mark= o,
    style=thick,
    dash pattern=on 10pt off 1pt
    ]
    coordinates {
    (IBM01,99.26)(IBM02,102.63)(IBM03,98.95)(IBM04,95.05)(IBM05,95.94)(IBM06,89.72)(IBM07,98.35)(IBM08,94.05)(IBM09,94.03)(IBM10,94.30)(IBM11,96.34)(IBM12,101.19)(IBM13,94.17)(IBM14,92.10)(IBM15,101.35)(IBM16,90.56)(IBM17,103.32)(IBM18,101.65)
};
\addlegendentry{KaHyPar}

\end{axis}
\end{tikzpicture}
}
\caption{Cut size with unit weight and $\epsilon = 1\%$, k = 4 . \protect\label{fig:fourpartition}}
\end{figure}

Table \ref{tab:table1} compares the cut sizes achieved by our HyperEF 2.0-based partitioner on the ISPD98 VLSI circuit benchmarks against hMETIS, SpecPart, KaHyPar, and MedPart. HyperEF 2.0 delivers average improvements of approximately 5\% at $\epsilon = 2\%$ and 4.82\% at $\epsilon = 10\%$ over hMETIS, and achieves gains of 0.69\% and 0.55\% over the best published results at the respective imbalance levels, highlighting its effectiveness. In several cases, it surpasses the best-known results by up to 5.5\%. To further evaluate generality, we applied our method to several instances of the Titan23 benchmark. As shown in Table \ref{tab:table_Titan23}, HyperEF 2.0 significantly outperforms hMETIS, achieving a 6.65\% improvement at $\epsilon = 2\%$ and a striking 29.38\% at $\epsilon = 20\%$, with the best average performance among all state-of-the-art methods. Furthermore, as shown in Fig. \ref{fig:fourpartition}, with four partitions and $\epsilon = 1\%$, HyperEF 2.0 consistently outperforms KaHyPar and hMETIS in cut size, normalized to hMETIS.
 
\begin{table*}[ht]
\caption{Cut size results of the ISPD98 VLSI circuit benchmark suite.}
\label{tab:table1}
\centering
\resizebox{.9\textwidth}{!}{
\footnotesize
\setlength{\tabcolsep}{2.85pt} 
\begin{tabular}{|c|c|c||c|c|c|c|c||c|c|c|c|c|}
\hline
Benchmark  & \multicolumn{2}{|c|}{Statistics} & \multicolumn{5}{|c|}{$\epsilon$ = $2\%$} & \multicolumn{5}{|c|}{$\epsilon$ = $10\%$}\\
\hline
   &$|V|$&$|E|$& SpecPart & hMETIS & KaHyPar & MedPart & HyperEF 2.0 & SpecPart & hMETIS & KaHyPar & MedPart & HyperEF 2.0\\
\hline
IBM01 & 12,752 & 14,111 & 202 & 213 & 202 & 202 & \textcolor{red}{\underline{201}} & 171 & 190 & 175 & \textcolor{red}{166} & \textcolor{red}{166}\\
\hline
IBM02 & 19,601 & 19,584 & 336 & 339 & 328 & 352 & \textcolor{red}{\underline{325}} & \textcolor{red}{262} & \textcolor{red}{262} & 263 & 264 & \textcolor{red}{262}\\
\hline
IBM03 & 23,136 & 27,401 & 959 & 972 & 958 & 955 & \textcolor{red}{\underline{952}} & 952 & 960 & \textcolor{red}{950} & 955 & \textcolor{red}{950}\\
\hline
IBM04 & 27,507 & 31,970 & 593 & 617 & \textcolor{red}{579} & 583 & \textcolor{red}{579} & \textcolor{red}{388} & \textcolor{red}{388} & \textcolor{red}{388} & 389 & \textcolor{red}{388}\\
\hline
IBM05 & 29,347 & 28,446 & 1720 & 1744 & 1713 & 1748 & \textcolor{red}{\underline{1707}} & 1688 & 1733 & 1698 & 1675 & \textcolor{red}{\underline{1645}}\\
\hline
IBM06 & 32,498 & 34,826 & \textcolor{red}{963} & 1037 & 978 & 1000 & \textcolor{red}{963} & \textcolor{red}{733} & 760 & 735 & 788 & \textcolor{red}{733}\\
\hline
IBM07 & 45,926 & 48,117 & 935 & 975 & 894 & 913 & \textcolor{red}{\underline{881}} & \textcolor{red}{760} & 796 & \textcolor{red}{760} & 773 & \textcolor{red}{760}\\
\hline
IBM08 & 51,309 & 50,513 & 1146 & 1146 & 1157 & 1158 & \textcolor{red}{\underline{1140}} & 1140 & 1145 & \textcolor{red}{1120} & 1131 & \textcolor{red}{1120}\\
\hline
IBM09 & 53,395 & 60,902 & \textcolor{red}{620} & 637 & \textcolor{red}{620} & 625 & \textcolor{red}{620} & \textcolor{red}{519} & 535 & \textcolor{red}{519} & 520 & \textcolor{red}{519}\\
\hline
IBM10 & 69,429 & 75,196 & 1318 & 1313 & 1339 & 1327 & \textcolor{red}{\underline{1254}} &1261 & 1284 & 1250 & 1259 & \textcolor{red}{\underline{1244}}\\
\hline
IBM11 & 70,558 & 81,454 & 1062 & 1114 & 1072 & 1069 & \textcolor{red}{\underline{1051}} & 764 & 782 & 769 & 774 & \textcolor{red}{\underline{763}}\\
\hline
IBM12 & 71,076 & 77,240 & \textcolor{red}{1920} & 1982 & 2163 & 1955 & \textcolor{red}{1920} & 1842 & 1940 & 1842 & 1914 & \textcolor{red}{\underline{1841}}\\
\hline
IBM13 & 84,199 & 99,666 & 848 & 871 & 848 & 850 & \textcolor{red}{\underline{831}} & 693 & 721 & 693 & 697 & \textcolor{red}{\underline{655}}\\
\hline
IBM14 & 147,605 & 152,772 & 1859 & 1967 & 1902 & 1876 & \textcolor{red}{\underline{1842}} & 1768 & 1665 & 1534 & 1639 & \textcolor{red}{\underline{1520}}\\
\hline
IBM15 & 161,570 & 186,608 & 2741 & 2886 & 2737 & 2896 & \textcolor{red}{\underline{2728}} & 2235 & 2262 & 2136 & 2169 & \textcolor{red}{\underline{2127}}\\
\hline
IBM16 & 183,484 & 190,048 & 1915 & 2095 & 1961 & 1972 & \textcolor{red}{\underline{1881}} & \textcolor{red}{1619} & 1708 & \textcolor{red}{1619} & 1645 & \textcolor{red}{1619}\\
\hline
IBM17 & 185,495 & 189,581 & 2354 & 2520 & \textcolor{red}{2284} & 2336 & 2285 & \textcolor{red}{1989} & 2300 & \textcolor{red}{1989} & 2024 & \textcolor{red}{1989}\\
\hline
IBM18 & 210,613 & 201,920 & 1535 & 1587 & 1949 & 1955 & \textcolor{red}{\underline{1521}} & 1537 & 1550 & 1915 & 1829 & \textcolor{red}{\underline{1520}}\\
\hline
\multicolumn{3}{|c|}{Average Improvement over hMETIS ($\%$)} & 3.64 & 0 & 1.68 & 1.11 & \textcolor{red}{5.03} & 2.91 & 0 & 2.52 & 1.65 & \textcolor{red}{4.82}\\
\hline
\end{tabular}
}
\end{table*}

\begin{table*}[ht]
\caption{Cut size results of the Titan23 benchmark suite.}
\label{tab:table_Titan23}
\centering
\resizebox{.9\textwidth}{!}{
\footnotesize
\setlength{\tabcolsep}{2.86pt} 
\begin{tabularx}{\textwidth}{|c|c|c||c|c|c|c|c||c|c|c|c|c|}
\hline
Benchmark  & \multicolumn{2}{|c|}{Statistics} & \multicolumn{5}{|c|}{$\epsilon$ = $2\%$} & \multicolumn{5}{|c|}{$\epsilon$ = $20\%$}\\
\hline
   &$|V|$&$|E|$& SpecPart & hMETIS & KaHyPar & MedPart & HyperEF 2.0 & SpecPart & hMETIS & KaHyPar & MedPart & HyperEF 2.0\\
\hline
sparcT1\_core & 91,976 & 92,827 & 1012 & 1066 & \textcolor{red}{974} & 1067 & \textcolor{red}{974} & 903 & 1290 & 873 & 624 & \textcolor{red}{\underline{583}}\\
\hline
neuron & 92,290 & 125,305 & 252 & 260 & 244 & 262 & \textcolor{red}{\underline{243}} & 206 & 270 & 244 & 270 & \textcolor{red}{\underline{200}}\\
\hline
stereo\_vision & 94,050 & 127,085 & 180 & 180 & \textcolor{red}{169} & 176 & \textcolor{red}{169} & \textcolor{red}{91} & 143 & \textcolor{red}{91} & 93 & \textcolor{red}{91}\\
\hline
des90 & 111,221 & 139,557 & 402 & 402 & 380 & \textcolor{red}{372} & 383 & 358 & 441 & 380 & \textcolor{red}{349} & 353\\
\hline
SLAM\_spheric & 113,115 & 142,408 & \textcolor{red}{1061} & \textcolor{red}{1061} & \textcolor{red}{1061} & \textcolor{red}{1061} & \textcolor{red}{1061} & \textcolor{red}{1061} & \textcolor{red}{1061} & \textcolor{red}{1061} & \textcolor{red}{1061} & \textcolor{red}{1061}\\
\hline
cholesky\_mc & 113,250 & 144,948 & 285 & 285 & \textcolor{red}{283} & \textcolor{red}{283} & \textcolor{red}{283} & 345 & 667 & 591 & \textcolor{red}{281} & \textcolor{red}{281}\\
\hline
segmemtation & 138,295 & 179,051 & 126 & 136 & \textcolor{red}{107} & 114 & \textcolor{red}{107} & \textcolor{red}{78} & 141 & \textcolor{red}{78} & \textcolor{red}{78} & \textcolor{red}{78}\\
\hline
bitonic\_mesh & 192,064 & 235,328 & \textcolor{red}{585} & 614 & 593 & 594 & \textcolor{red}{585} & \textcolor{red}{483} & 590 & 592 & 493 & 506\\
\hline
dart & 202,354 & 223,301 & 807 & 844 & 924 & 805 & \textcolor{red}{\underline{784}} & 540 & 603 & 594 & 549 & \textcolor{red}{\underline{539}}\\
\hline
\multicolumn{3}{|c|}{Average Improvement over hMETIS ($\%$)} & 2.73 & 0 & 4.70 & 3.73 & \textcolor{red}{6.65} & 25.59 & 0 & 16.59 & 26.15 & \textcolor{red}{29.38}\\
\hline

\end{tabularx}
}
\end{table*}

\subsubsection{Ablation Study: Clustering Quality and Efficiency}
\begin{figure}[tbp]
    \centering
    \resizebox{0.6\linewidth}{!}{
\begin{tikzpicture}
\begin{axis}[
    ylabel={Runtime (s)},
    symbolic x coords={IBM01, IBM02, IBM03, IBM04, IBM05, IBM06, IBM07, IBM08, IBM09, IBM10, IBM11, IBM12, IBM13, IBM14, IBM15, IBM16, IBM17, IBM18},
    xtick=data,
    x tick label style={rotate=90, anchor=east, font=\scriptsize}, 
    ytick={0,10,...,150},
    ymin=-5,
    ymax=150,
    legend pos = north west,
    ymajorgrids=false,
    xmajorgrids=true,
    grid style=dashed,
]

\addplot[
    color=blue,
    style=thick,
    dash pattern=on 10pt off 1pt
    ]
    coordinates {
    (IBM01,9.12)(IBM02,9.95)(IBM03,10.30)(IBM04,10.69)(IBM05,11.14)(IBM06,11.20)(IBM07,11.56)(IBM08,12.64)(IBM09,12.51)(IBM10,14.58)(IBM11,14.14)(IBM12,14.64)(IBM13,16.02)(IBM14,22.08)(IBM15,28.24)(IBM16,30.49)(IBM17,31.48)(IBM18,33.28)
};
\addlegendentry{HyperEF 2.0}
\addplot[
    color=red,
    style=thick,
    dash pattern=on 10pt off 1pt
    ]
    coordinates {
    (IBM01,13.88)(IBM02,19.12)(IBM03,17.63)(IBM04,17.75)(IBM05,25.88)(IBM06,23.19)(IBM07,23.31)(IBM08,47.76)(IBM09,24.45)(IBM10,32.93)(IBM11,27.49)(IBM12,40.30)(IBM13,41.83)(IBM14,59.64)(IBM15,103.34)(IBM16,85.86)(IBM17,116.41)(IBM18,140.00)
};
\addlegendentry{HyperSF}

\end{axis}
\end{tikzpicture}
}
\caption{Run time analysis for ISPD98 benchmarks. \protect\label{fig:ISPDchart}}
\end{figure}
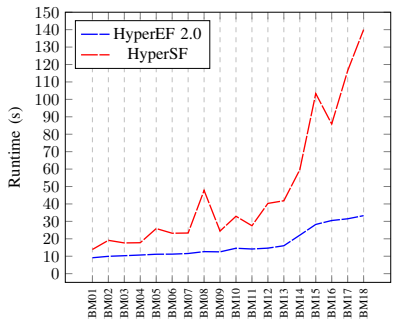

To evaluate the quality of our proposed spectral clustering method, we compare it directly with HyperEF and HyperSF \cite{aghdaei2021hypersf} in the context of community detection (CD). Specifically, we configure KaHyPar to use only its community detection component and replace its default Louvain-based method with (1) our HyperEF 2.0 clustering, (2) HyperEF’s clustering, and (3) HyperSF’s clustering. Fig. \ref{fig:cutsize} presents normalized (to Louvain) cut sizes across several benchmarks at $k =2, \epsilon = 2\%$, where HyperEF 2.0 consistently achieves lower cut sizes, demonstrating superior clustering quality. Additionally, runtime comparisons in Fig. \ref{fig:ISPDchart} show that our method is up to $4.5\times$ faster than HyperSF, highlighting both its effectiveness and computational efficiency.

\begin{figure}[tbp]
    \centering
    \resizebox{0.6\linewidth}{!}{
\begin{tikzpicture}
\begin{axis}[
    ylabel={Normalized cut size},
    symbolic x coords={IBM01, IBM02, IBM03, IBM04, IBM05, IBM06, IBM07, IBM08, IBM09, IBM10, IBM11, IBM12, IBM13, IBM14, IBM15, IBM16, IBM17, IBM18},
    xtick=data,
    x tick label style={rotate=90, anchor=east, font=\scriptsize}, 
    ytick={70,75,...,110},
    ymin=75,
    ymax=115,
    legend pos = south west,
    ymajorgrids=false,
    xmajorgrids=true,
    grid style=dashed,
]

\addplot[
    color=blue,
    mark=star,
    style=thick,
    dash pattern=on 10pt off 1pt
    ]
    coordinates {
    (IBM01,100)(IBM02,99.08)(IBM03,99.37)(IBM04,100.17)(IBM05,99.76)(IBM06,99.38)(IBM07,98.54)(IBM08,98.53)(IBM09,100)(IBM10,93.65)(IBM11,98.04)(IBM12,91.77)(IBM13,97.99)(IBM14,97.21)(IBM15,100.14)(IBM16,103.92)(IBM17,100.43)(IBM18,78.04)
};
\addlegendentry{HyperEF 2.0 CD}

\addplot[
    color=red,
    mark= o,
    style=thick,
    dash pattern=on 10pt off 1pt
    ]
    coordinates {
    (IBM01,100)(IBM02,100)(IBM03,99.37)(IBM04,103.11)(IBM05,100.35)(IBM06,101.12)(IBM07,100.89)(IBM08,98.53)(IBM09,100)(IBM10,98.05)(IBM11,111.28)(IBM12,91.86)(IBM13,97.99)(IBM14,97.21)(IBM15,107.78)(IBM16,109.02)(IBM17,100.43)(IBM18,78.04)
};
\addlegendentry{HyperSF CD}

\addplot[
    color=black,
    mark= -.,
    style=thick,
    dash pattern=on 10pt off 1pt
    ]
    coordinates {
    (IBM01,100)(IBM02,100.3)(IBM03,102.4)(IBM04,104.67)(IBM05,99.77)(IBM06,100.3)(IBM07,104.5)(IBM08,98.53)(IBM09,102.26)(IBM10,98.43)(IBM11,111.28)(IBM12,91.86)(IBM13,98)(IBM14,97.21)(IBM15,103.10)(IBM16,109.02)(IBM17,100.43)(IBM18,78.04)
};
\addlegendentry{HyperEF CD}

\end{axis}
\end{tikzpicture}
}
\caption{Cut size improvement with spectral clustering. \protect\label{fig:cutsize}}
\end{figure}
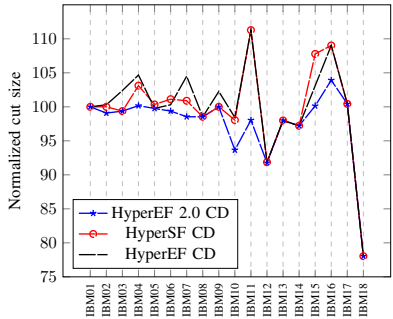

\section{Conclusion}\label{sec:conclusion}
In this study, we introduced HyperEF 2.0, an advanced hypergraph coarsening framework that outperforms previous approaches in both performance and runtime efficiency. Our method builds upon the concept of hypergraph effective resistance, enhancing its estimation through an improved Krylov subspace technique, and incorporates a novel resistance-based local clustering algorithm to enhance cluster quality. We further integrate this framework into hypergraph partitioning tasks. Extensive experiments on real-world VLSI benchmarks demonstrate that HyperEF 2.0 consistently achieves lower cluster conductance and significantly accelerates computation. Moreover, it results in a substantial reduction in partitioning cut size compared to state-of-the-art techniques, validating the effectiveness of our coarsening strategy in preserving the essential structure of the hypergraph.

\section*{Acknowledgment}
This work is supported in part by  the National Science Foundation under Grants    CCF-2417619, CCF-2021309,  CCF-2011412, and CCF-2212370.

\bibliographystyle{ieeetr}
\bibliography{hyperef_2}

\end{document}